\documentclass[aps,pra,twocolumn,showpacs,superscriptaddress]{revtex4-1}
\usepackage{bm}
\usepackage{mathrsfs}
\usepackage{amsmath}
\usepackage{amssymb}
\usepackage{graphicx}
\usepackage{amsfonts}
\usepackage{amsthm}
\usepackage{color}
\usepackage{dcolumn}
\usepackage{txfonts}

\begin{document}

\title{Fidelity Susceptibility in the Quantum Rabi Model}
\author{Bo-Bo Wei}
\affiliation{School of Physics and Energy, Shenzhen University, Shenzhen 518060, China}
\email{Corresponding author: bbwei@szu.edu.cn}
\author{Xiao-Chen Lv}
\affiliation{School of Physics and Energy, Shenzhen University, Shenzhen 518060, China}

\begin{abstract}
Quantum criticality usually occurs in many-body systems. Recently it was shown that the quantum Rabi model, which describes a two-level atom coupled to a single model cavity field, presents quantum phase transitions from a normal phase to a superradiate phase when the ratio between the frequency of the two level atom and the frequency of the cavity field extends to infinity. In this work, we study quantum phase transitions in the quantum Rabi model from the fidelity susceptibility perspective. We found that the fidelity susceptibility and the generalized adiabatic susceptibility present universal finite size scaling behaviors near the quantum critical point of the Rabi model if the ratio between frequency of the two level atom and frequency of the cavity field is finite. From the finite size scaling analysis of the fidelity susceptibility, we found that the adiabatic dimension of the fidelity susceptibility  and the generalized adiabatic susceptibility of fourth order in the Rabi model are $4/3$ and $2$, respectively. Meanwhile the correlation length critical exponent and the dynamical critical exponent in the quantum critical point of the Rabi model are found to be $3/2$ and $1/3$ respectively. Since the fidelity susceptibility and the generalized adiabatic susceptibility are the moments of the quantum noise spectrum which is directly measurable by experiments in linear response regime, the scaling behavior of the fidelity susceptibility in the Rabi model could be tested experimentally. The simple structure of the quantum Rabi model paves the way for experimentally observing the universal scaling behavior of the fidelity susceptibility at a quantum phase transition.
\end{abstract}


\maketitle

\section{Introduction}
Quantum criticality is one of the most intriguing phenomena since it signals emergence of new states of matter \cite{QPT2011}. However, to observe exotic features at quantum critical point, one has to study systems in the thermodynamic limit involving a large number of system components, which put obstacles both experimentally and theoretically \cite{Cardy1996}. Recently, it was found that \cite{RabiQPT2015} the quantum Rabi model which consists of a single model cavity field and a two level atom, presents quantum phase transitions (QPT) from a normal phase to a superradiant phase as the ratio $\eta\equiv\Omega/\omega_0$ between the frequency of the two level atom $\Omega$ and the frequency of the single mode cavity $\omega_0$ extends to infinity. It has long been known that the Dicke model, many atoms generalization of the Rabi model, presents a QPT as the number of atoms $N\rightarrow\infty$. Hwang and Plenio \cite{RabiQPT2015} demonstrate that the Rabi model itself undergoes a QPT with the same universal properties as that of the Dicke model when the ratio $\eta$ between the transition frequency of the two level atom and the cavity frequency diverges. At the same time the finite-frequency scaling exponents for $\eta$ are identical to those for $N$. QPT in the Rabi model opens up a new paradigm for studying phase transitions in the photonic systems \cite{RabiQPT2015,JCQPT2016}.

In recent decades, there is a great deal of interest in studying quantum
phase transitions from the perspective of quantum information science \cite{QI2000}, such as from perspective of
quantum entanglement \cite{entanglementQPT2002,entanglementQPT2008} and viewpoint of quantum fidelity \cite{Sun2006,Zanardi2006,GuReview}. The advantage of studying quantum phase transitions using concepts in quantum information science is that no prior knowledge of the order parameter and the symmetry of the system are required \cite{GuReview,You2007,Zanardi2007,Venuti2007,YangMF2007,YangMF2008,Paun2008,Chen2008,Gu2008a,Gu2008b,Gu2008c,Gu2009,Schwandt2009,fscaling2010,fs2011,fs2012a,fs2012b,fs2012c,fs2013a,fs2013b,fs2013c,fs2014a,fs2014b,fs2015a,fs2015b,fs2017}. These works bridge the quantum information theory and condensed matter
physics and shall deepen our understanding on the various condensed matter phenomena.

The quantum fidelity approach to quantum phase transitions possess a clearer physical picture\cite{GuReview}. It is generally believed that the
ground state wave functions at two sides of the critical point $g_c$ of a quantum many body system have qualitatively different structures and thus the overlap of two
ground states separated by a small distance $\delta g$ in the parameter space, i.e. $|\langle\Psi_0(g)|\Psi_0(g+\delta g)\rangle|$
presents a minimum at the critical point $g_c$. Because the structure of the ground state
of a quantum many body system experiences a significant change as the
system is driven adiabatically across the transition point,
we can also imagine that the leading term of the fidelity,
i.e. the fidelity susceptibility which denotes the leading
response of the ground state to the driving parameter,
should be a maximum or even divergent at the transition
point \cite{You2007}. Quantum phase transitions from perspective
of the fidelity and fidelity susceptibility have been investigated in many different systems \cite{GuReview,You2007,Zanardi2007,Venuti2007,YangMF2007,YangMF2008,Paun2008,Chen2008,Gu2008a,Gu2008b,Gu2008c,Gu2009,Schwandt2009,fscaling2010,fs2011,fs2012a,fs2012b,fs2012c,fs2013a,fs2013b,fs2013c,fs2014a,fs2014b,fs2015a,fs2015b,fs2017} and it was shown that fidelity susceptibility providers a simple approach to determining the universality of quantum phase transitions \cite{Zanardi2007,Venuti2007,YangMF2007,YangMF2008,Paun2008,Chen2008,Gu2008a,Gu2008b,Gu2008c,Gu2009,Schwandt2009,fscaling2010,fs2011,fs2012a,fs2012b,fs2012c,fs2013a,fs2013b,fs2013c,fs2014a,fs2014b,fs2015a,fs2015b,fs2017}.

In this work, we study the quantum phase transitions in the Rabi model from the perspective of fidelity susceptibility. Having only two constituent particles, the Rabi model is far
from being in the thermodynamic limit where a QPT typically occurs; however, a ratio of the atomic transition frequency to the cavity field frequency that approaches infinity, can play the role of a thermodynamic
limit. We investigated the universal finite size scaling behavior of fidelity susceptibility and the generalized adiabatic susceptibility in the Rabi model when the ratio $\eta$ between the frequency of the two level atom and the frequency of the cavity field is finite. From the universal scaling analysis of the fidelity susceptibility and the generalized adiabatic susceptibility in the Rabi model, we found that the adiabatic dimension of the fidelity susceptibility and the generalized adiabatic susceptibility of fourth order in the Rabi model are $4/3$ and $2$, respectively and the critical exponents for the correlation length and the dynamical critical exponent are respectively $\nu=3/2$ and $z=1/3$. Because the simplicity of the structure of the Rabi model where the ratio $\eta$ plays the role of particle number, our results for the universal finite size scaling behavior of fidelity susceptibility in the Rabi model could be verified experimentally.

This paper is structured as follows. In Sec.~II, we review briefly the quantum Rabi model and its quantum phase transitions. In Sec.~III, we introduce the concept of fidelity susceptibility and the generalized adiabatic susceptibility in general situations. Sec.~IV is devoted to study the universal scaling behavior of the fidelity susceptibility in the quantum Rabi model and in Sec.~V, we investigate the scaling behavior of the generalized adiabatic susceptibility in the Rabi model. In Sec.~VI, we discuss the possible experimental detection of the fidelity susceptibility and generalized adiabatic susceptibility. Finally Sec.~VII is a brief summary.

\begin{figure}
\begin{center}
\includegraphics[width=\columnwidth]{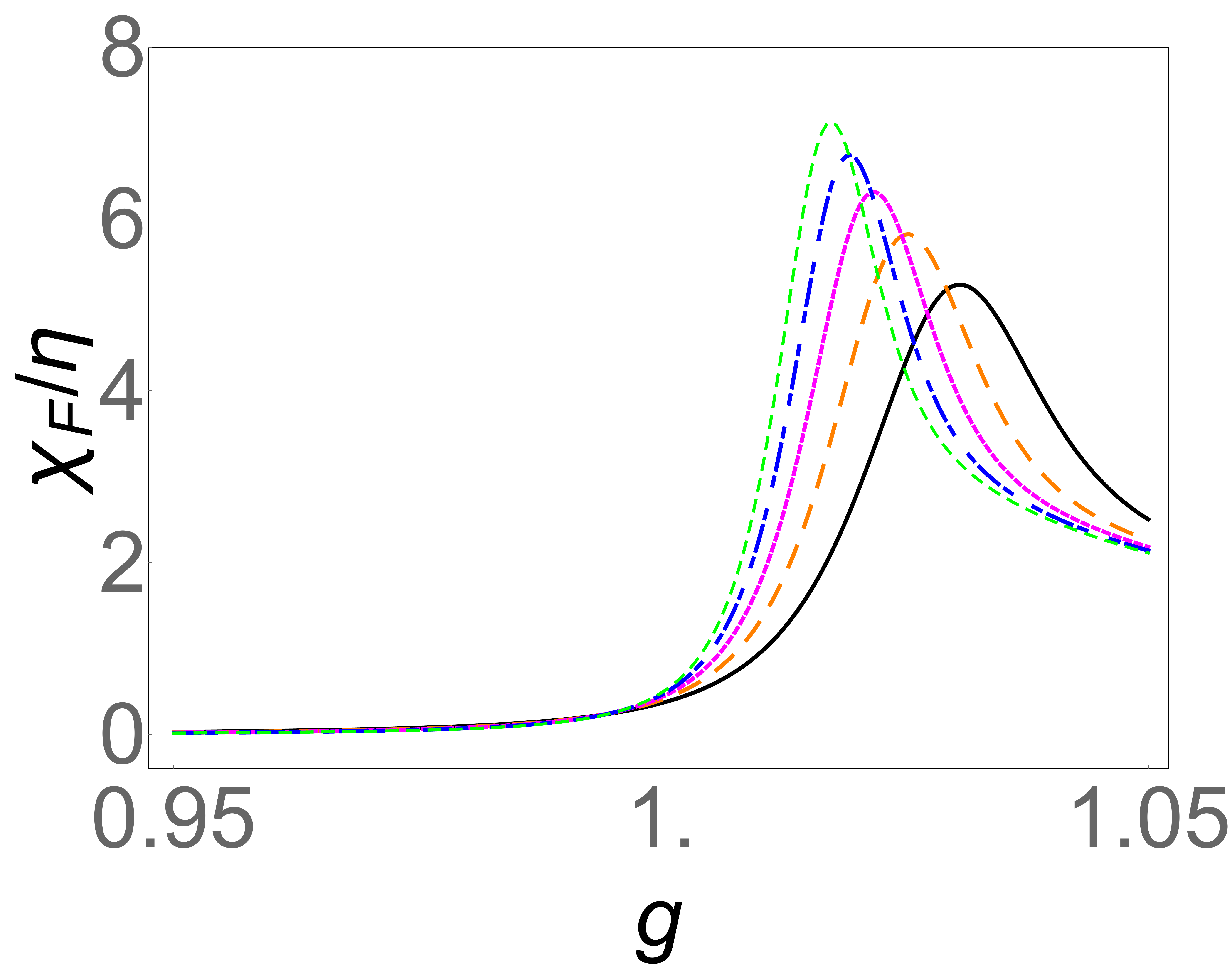}
\end{center}
\caption{(color online). Fidelity susceptibility, $\chi_F$, in the Rabi model as a function of control parameter $g$ for different $\eta$ which is the ratio between the frequency of the two level atom and the frequency of the cavity field. The black solid line is $\eta=300$, the orange dashed line is $\eta=400$, the magenta dotted line is $\eta=500$, the blue dash-dotted line is $\eta=600$ and the purple short dashed line is $\eta=700$.}
\label{fig:epsart1}
\end{figure}

\section{Quantum Phase Transitions in the Rabi Model.}
\noindent The Hamiltonian of the quantum Rabi model is
\begin{eqnarray}
H_{\text{Rabi}}&=&\omega_0a^{\dagger}a+\frac{\Omega}{2}\sigma_z-\lambda(a+a^{\dagger})\sigma_x.
\end{eqnarray}
Here $\sigma_x$ and $\sigma_z$ are Pauli matrices for a two-level atom and $a$ and $a^{\dagger}$ are respectively the annihilation and creation operator of a cavity field.
The cavity field frequency is $\omega_0$ and the transition frequency of the two level atom is $\Omega$. The coupling strength between the cavity field and the two level atom is $\lambda$.
Defining dimensionless parameters $\eta=\Omega/\omega_0$ and $g=2\lambda/\sqrt{\Omega\omega_0}$, the Rabi Hamiltonian can be rewritten as
\begin{eqnarray}
H_{\text{Rabi}}/\omega_0&=&a^{\dagger}a+\frac{\eta}{2}\sigma_z-\frac{g\sqrt{\eta}}{2}(a+a^{\dagger})\sigma_x,\\
&\equiv&H_0+gH_1.
\end{eqnarray}
Recently it was shown that this Rabi model presents second order quantum phase transition from a normal phase to a superradiant phase at the critical point $g=1$ in the $\eta\rightarrow\infty$ \cite{RabiQPT2015}. For large but not infinite $\eta$, the ground state
expectation values and the energy spectrum exhibit a
critical scaling in $\eta$ at $g=1$, so-called finite-frequency
scaling \cite{RabiQPT2015}.

\begin{figure}
\begin{center}
\includegraphics[width=\columnwidth]{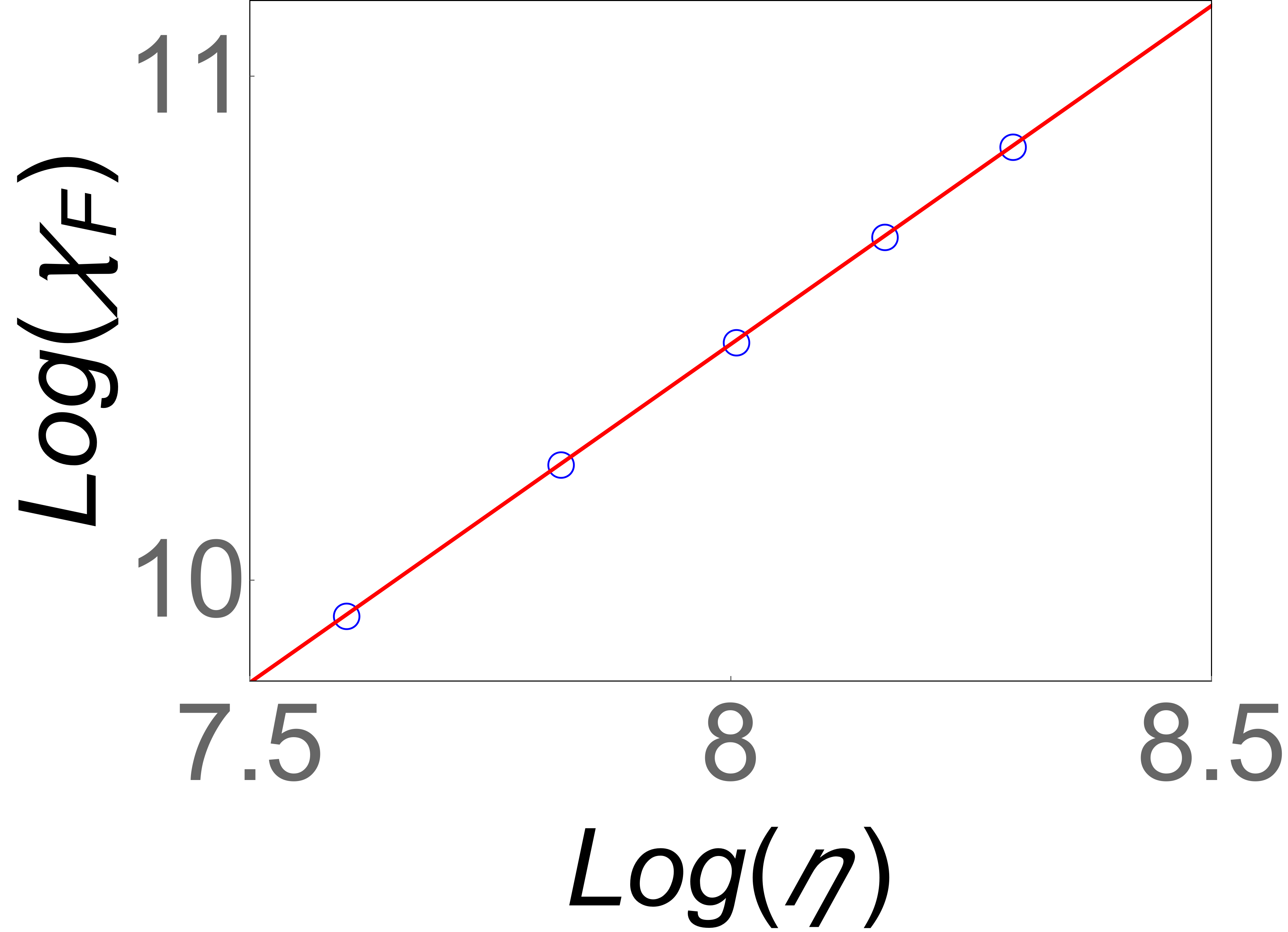}
\end{center}
\caption{(color online). The finite size scaling of the amplitude of fidelity susceptibility at the peak position $\ln[\chi_F(g_m)]$ as a function of $\ln (\eta)$. The blue hollow circle is the exact numerical result and the red line is a linear fit of the numerical data. From the linear fit, we find that the adiabatic dimension of the fidelity susceptibility, $\chi_F\sim \eta^{\mu}$, i.e. $\mu=1.34$.}
\label{fig:epsart2}
\end{figure}

\section{Fidelity Susceptibility and the generalized adiabatic susceptibility}
Let us consider a general many body system with Hamiltonian
\begin{eqnarray}
H(g)=H_0+gH_1,
\end{eqnarray}
where $H_1$ is the driven Hamiltonian which drives the quantum phase transitions from one phase to the other with strength $g$. The quantum fidelity of the ground state \cite{Sun2006,Zanardi2006,GuReview} is defined as the overlap between two ground states $|\Psi_0(g)\rangle$ and $|\Psi_0(g+\delta g)\rangle$,
\begin{eqnarray}
F(g,\delta g)=\left|\langle\Psi_0(g)|\Psi_0(g+\delta g)\right|.
\end{eqnarray}
Because the ground state $|\Psi_0(g)\rangle$ is qualitatively different in two sides of a quantum critical point, the fidelity could characterize quantum phase transitions by presenting a sharp dip near the critical point. It is clear that fidelity depends on two parameters $g$ and $\delta g$. The leading contributions to the fidelity is the fidelity susceptibility \cite{You2007}, which is defined in the limit of $\delta g\rightarrow0$,
\begin{eqnarray}
\chi_F(g)=\lim_{\delta g\rightarrow0}\frac{-2\ln F(g,\delta g)}{(\delta g)^2}.
\end{eqnarray}
In terms of the eigen states of the Hamiltonian, the fidelity susceptibility is given by \cite{You2007}
\begin{eqnarray}\label{fs}
\chi_F(g)=\sum_{n\neq0}\frac{\left|\langle\Psi_n(g)|H_1|\Psi_0(g)\right|^2}{[E_n(g)-E_n(0)]^2}.
\end{eqnarray}
Here $|\Psi_n(g)\rangle$ is the eigen states of $H(g)$ with corresponding eigen energy $E_n(g)$, i.e. $H(g)|\Psi_n(g)\rangle=E_n(g)|\Psi_n(g)\rangle$.

Many works have been devoted to investigate the relationship between fidelity susceptibility and quantum phase transitions \cite{You2007,Zanardi2007,Venuti2007,YangMF2007,YangMF2008,Paun2008,Chen2008,Gu2008a,Gu2008b,Gu2008c,Gu2009,Schwandt2009,fscaling2010,fs2011,fs2012a,fs2012b,fs2012c,fs2013a,fs2013b,fs2013c,fs2014a,fs2014b,fs2015a,fs2015b,fs2017}. It is shown that the finite size scaling of the fidelity susceptibility in the neighborhood of a quantum critical point is \cite{GuReview}
\begin{eqnarray}
\chi_F(g_m)\propto L^{\mu},
\end{eqnarray}
where $L$ is the size of the system and $\mu$ is the critical adiabatic dimension and $g_m$ is the position of the parameter where the fidelity susceptibility is maximum. For finite size system, the fidelity susceptibility presents scaling behavior as
\cite{Gu2008a,fscaling2010}
\begin{eqnarray}
\chi_F(g)=L^{2/\nu}f((g-g_m)L^{1/\nu}).
\end{eqnarray}
Here $\nu$ is the critical exponent of the correlation length and $f(x)$ is an unknown but universal function in the sense that it is independent of the size of the system. Thus the investigation of fidelity susceptibility provides a simple approach to determine the universality class of a quantum phase transitions by extracting the universal critical exponents $\mu$ and $\nu$ \cite{GuReview}.

Consider a time dependent fidelity $F(t)=\left|\langle\Psi(t)|\Psi_0(t)\rangle\right|$. Here $|\Psi_0(t)\rangle$ is the instantaneous ground state and $|\Psi(t)\rangle$ is a time dependent state because of time-dependent driven. When the tuning parameter $g$ is time dependent, i.e. $g(t)=g_c+bt^r/r!\theta(t)$ with $\theta(t)$ being the step function and $b$ is the adiabatic parameter that controls the distance
to the critical point. Because the energy gap at the critical point of QPT $g_c$ tends to zero when the size of the system extends to infinity, in such a case it is impossible to pass through the critical point
at a finite speed $b$ without exciting the system. The probability of excitations is given by \cite{generalized2010a,generalized2010b,noneqreview2011}
\begin{eqnarray}
P_{\text{ex}}&=&1-F(t)^2=b^2\chi_{2r+2}(g_c).
\end{eqnarray}
Here
\begin{eqnarray}\label{gs}
\chi_{2r+2}(g)&=&\sum_{n\neq0}\frac{\left|\langle\Psi_n(g)|H_1|\Psi_0(g)\right|^2}{[E_n(g)-E_n(0)]^{2r+2}},
\end{eqnarray}
is the generalized adiabatic susceptibility of order $2r+2$ \cite{generalized2010a,generalized2010b,noneqreview2011}. The fidelity susceptibility is the generalized adiabatic susceptibility with $r=0$. For $r=1$, we have the generalized adiabatic susceptibility of order four, $\chi_4$. It is demonstrated that the finite size scaling of the generalized adiabatic susceptibility in the neighborhood of a quantum critical point is \cite{noneqreview2011}
\begin{eqnarray}
\chi_{2r+2}(g_m)\propto L^{2/\nu+2zr},
\end{eqnarray}
where $g_m$ is the position of the parameter where the fidelity susceptibility is maximum and the critical adiabatic dimension of the generalized adiabatic susceptibility of order $2r+2$ is $2/\nu+2zr$ with $z$ being the dynamical critical exponent. For finite size system, the generalized adiabatic susceptibility presents scaling behavior as
\cite{noneqreview2011}
\begin{eqnarray}
\chi_{2r+2}(g)=L^{2/\nu+2zr}f_r((g-g_m)L^{1/\nu}).
\end{eqnarray}
Here $f_r(x)$ is the universal scaling function of the generalized adiabatic susceptibility $\chi_{2r+2}$. Investigation of fidelity susceptibility and the generalized fidelity susceptibility provide a simple approach to determine the universality class of a quantum phase transitions by extracting the universal critical exponents $\mu,\nu,z$ \cite{GuReview}.

\begin{figure}
\begin{center}
\includegraphics[width=\columnwidth]{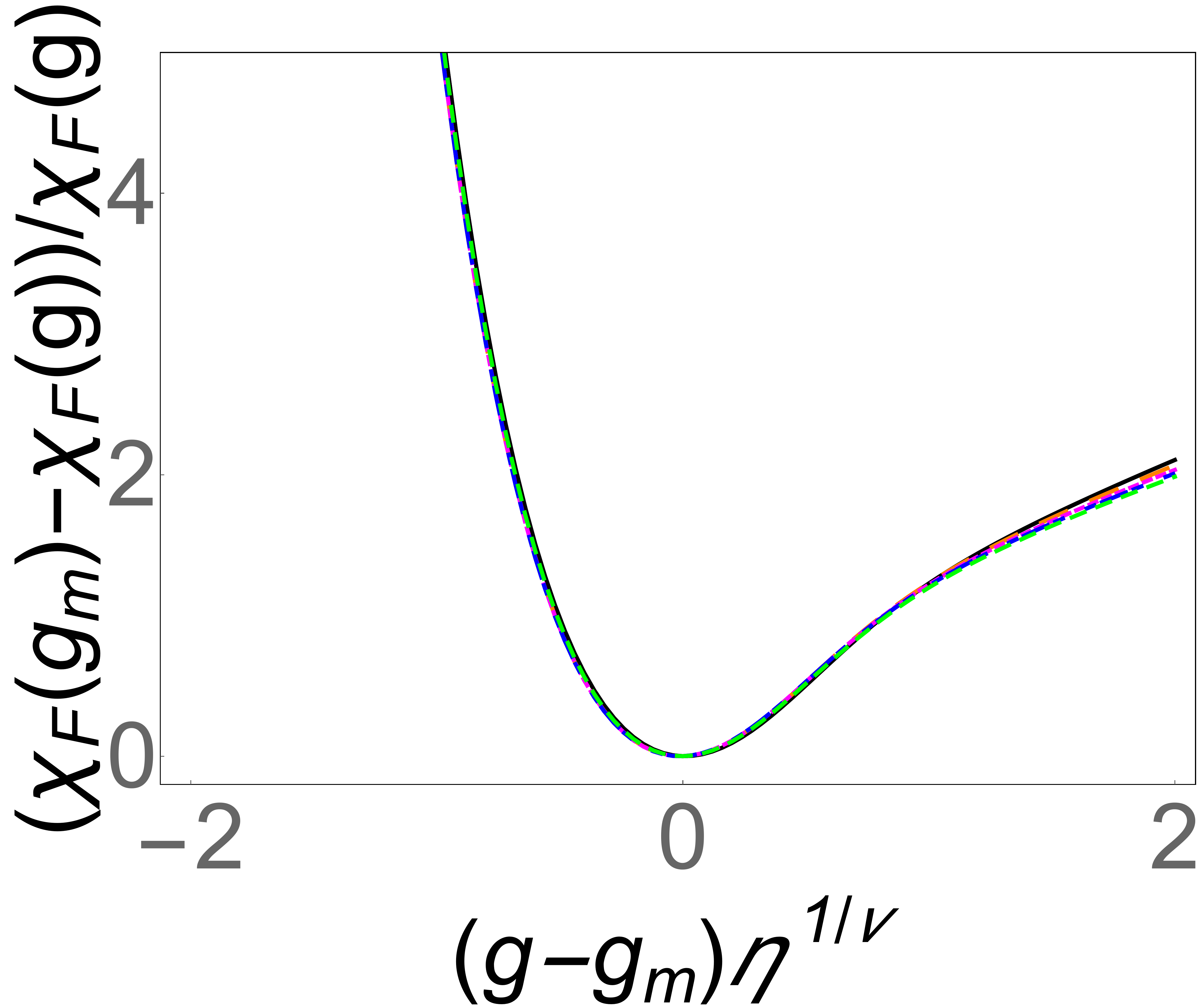}
\end{center}
\caption{(color online). Data collapse of fidelity susceptibility in the Rabi model shown in Figure 1. $\chi_F$ is a function of $\eta^{1/\nu}(g-g_m)$ only with $g_m$ being the position of the maximum of the fidelity susceptibility and $\nu=1.49$ being chosen so that data for different $\eta$ collapse perfectly.}
\label{fig:epsart3}
\end{figure}

\section{Fidelity susceptibility in the Rabi model}

The Hamiltonian of the Rabi model can be numerically diagonalized when we truncate the Hilbert space by constraining the maximum number of photons $N=a^{\dagger}a$. Then the fidelity susceptibility can also be exactly evaluated by Equation \eqref{fs}. We try different number of photons in truncating the Hilbert space until we get convergent result for the fidelity susceptibility. We show the numerical exact results for the fidelity susceptibility in the Rabi model as a function of control parameter $g$ for different $\eta=300,400,500,600,700$ respectively in Figure 1. First of all, one can see that the fidelity susceptibility present a peak at $g=g_m$ around the critical point of the Rabi model at $g_c=1$. Note that the position of parameter where fidelity susceptibility is maximum is different from $g_c$ is a finite size effect.  Second, when $\eta$ increases, the peak structure in the fidelity susceptibility becomes sharper and higher and the position of the control parameter $g_m$ where the fidelity susceptibility is maximum approach to the quantum critical point of the Rabi model $g_c=1$. Third, the ratio $\eta$ between the frequency of the two level atom and the frequency of the cavity field plays the same role at that of the number of particles in ordinary phase transitions.

We further study the adiabatic dimension of the fidelity susceptibility in the Rabi model in Figure 2. The blue hollow circle is the numerical results of the peak value of the fidelity susceptibility $\log(\chi_F)$ for different $\eta$. We make a linear fit of these data and find the straight line being, $-0.27+1.34x$ (see the red line in Figure 2). Thus the critical adiabatic dimension of the fidelity susceptibility, $\chi_F\sim \eta^{\mu}$, is $\mu=1.34$, which is close to the exact result $\mu=2/\nu=4/3$ \cite{RabiQPT2015}.

Figure 3 shows the data collapse of the fidelity susceptibility in the Rabi model for different finite ratio $\eta$. Finite size scaling shows that the fidelity susceptibility is a universal function of $\eta^{1/\nu}(g-g_m)$ with $g_m$ being the position of the control parameter where the fidelity susceptibility is maximum. The correlation length critical exponent $\nu$ is chosen in order that the data for different $\eta$ collapse perfectly and we found that $\nu=1.49$, which agrees with the exact result of the correlation length exponent $\nu=3/2$ of the Rabi model \cite{RabiQPT2015}.

\begin{figure}
\begin{center}
\includegraphics[width=\columnwidth]{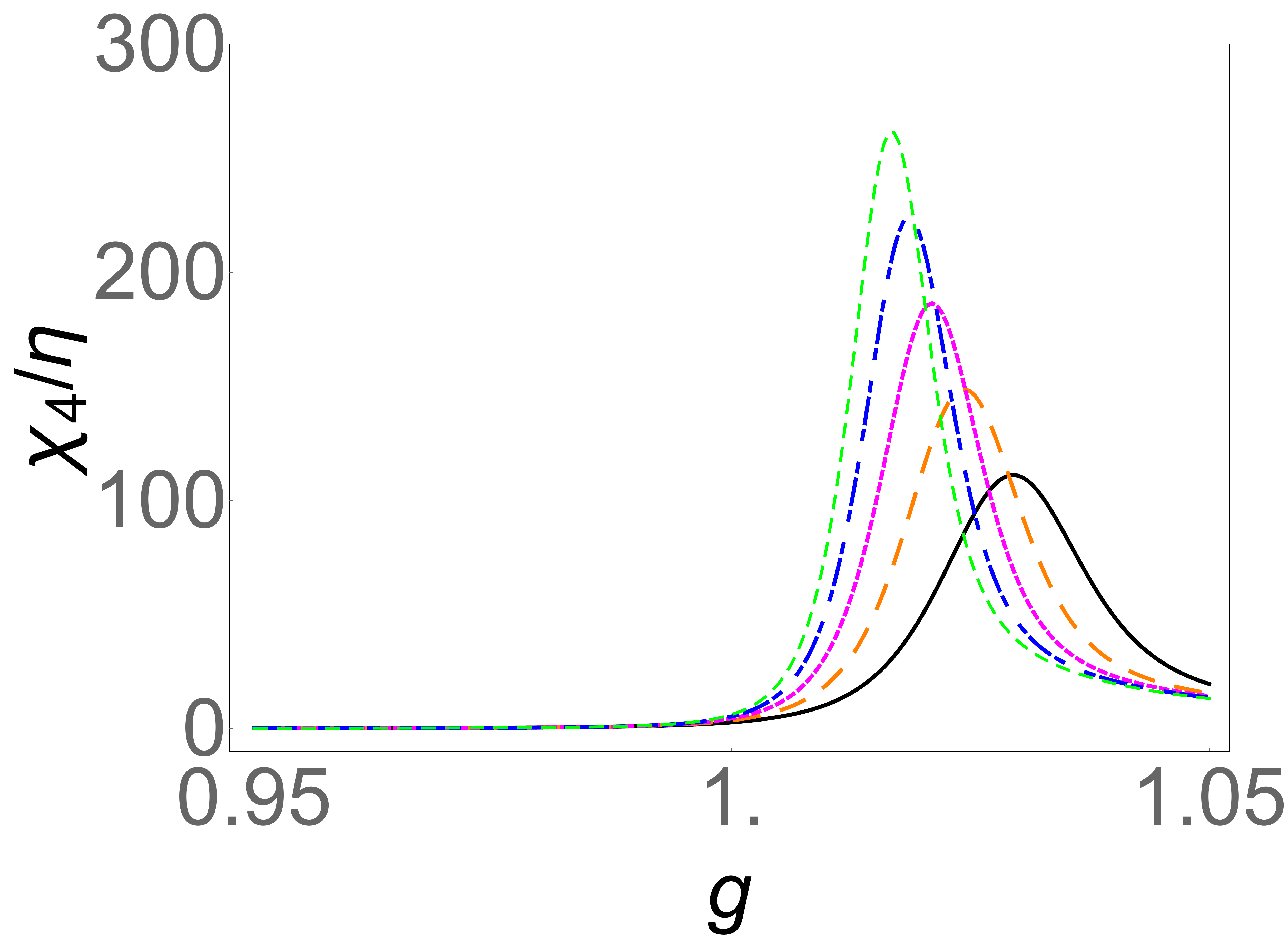}
\end{center}
\caption{(color online). Generalized fidelity susceptibility, $\chi_4$, in the Rabi model as a function of control parameter $g$ for different $\eta$. The black solid line is $\eta=300$, the orange dashed line is $\eta=400$, the magenta dotted line is $\eta=500$, the blue dash-dotted line is $\eta=600$ and the purple short dashed line is $\eta=700$.}
\label{fig:epsart4}
\end{figure}

\begin{figure}
\begin{center}
\includegraphics[width=\columnwidth]{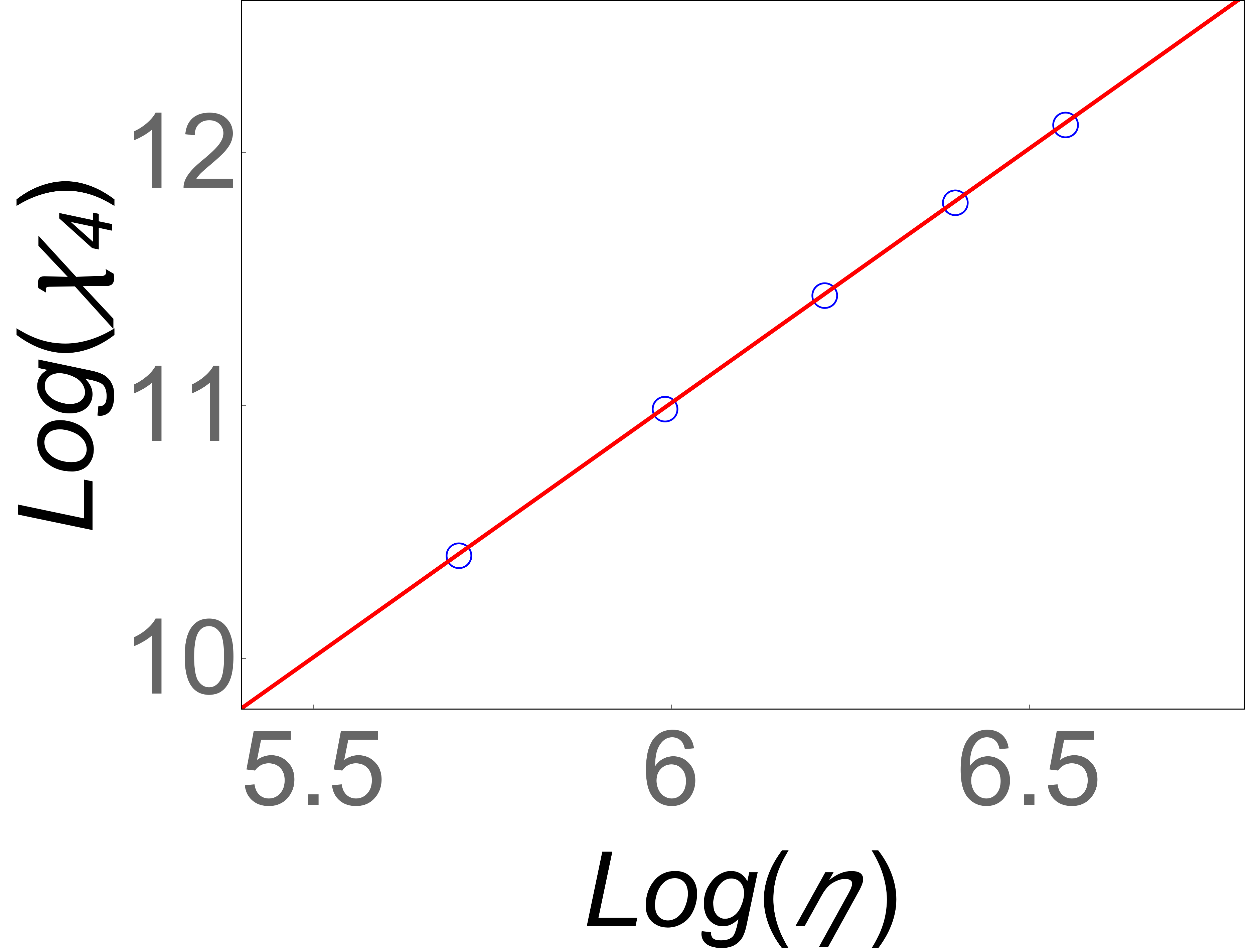}
\end{center}
\caption{(color online). The finite size scaling of the amplitude of generalized fidelity susceptibility $\chi_4$ at the peak position $\ln[\chi_4(g_m)]$ as a function of $\ln (\eta)$. The blue hollow circle is the exact numerical result and the red line comes from a linear fit of the numerical data. From the linear fit, we find that the adiabatic dimension of the generalized fidelity susceptibility, $\chi_4\sim \eta^{\mu}$, i.e. $\mu=2.01$.}
\label{fig:epsart5}
\end{figure}

\section{Generalized adiabatic susceptibility in the Rabi model}

We show the numerical exact results for the generalized adiabatic susceptibility $\chi_4$ in the Rabi model as a function of control parameter $g$ for different $\eta=300,400,500,600,700$ respectively in Figure 4. One can see that the generalized adiabatic susceptibility has a peak around the critical point of the Rabi model at $g_c=1$. Secondly, when $\eta$ increases, the peak structure in the generalized adiabatic susceptibility becomes sharper and higher and the position of the control parameter $g_m$ where the generalized adiabatic susceptibility is maximum is closer to the critical point $g_c=1$ of the Rabi model.

We further study the adiabatic dimension of the generalized adiabatic susceptibility $\chi_4$ in the Rabi model in Figure 5. The blue hollow circle is the numerical results of the peak of the generalized adiabatic susceptibility $\log(\chi_4)$ for different $\eta$. We make a linear fit of these data with the straight line being, $-1.05+2.01x$ (red line in Figure 5). Thus the critical adiabatic dimension of the generalized adiabatic susceptibility, $\chi_4\sim \eta^{\mu}$, is $\mu=2.01$, which agrees with the exact results $\mu=2/\nu+2z=2$ \cite{RabiQPT2015}. From the adiabatic dimension of the generalized fidelity susceptibility, we found that the dynamical critical exponent in the critical point of the Rabi model is $z\approx0.34$, which is close to the exact result $z=1/3$ \cite{RabiQPT2015}.

Figure 6 shows the data collapse of the generalized adiabatic susceptibility in the Rabi model for different ratio $\eta$. Finite size scaling theory tells us that the generalized adiabatic susceptibility is a universal function of $\eta^{1/\nu}(g-g_m)$ with $g_m$ being the position of the control parameter where the generalized adiabatic susceptibility is maximum. The correlation length critical exponent $\nu$ is chosen in order that the data of the generalized adiabatic susceptibility $\chi_4$ for different $\eta$ collapse perfectly and we found that $\nu=1.49$, which agrees with the exact result of the correlation length exponent $\nu=3/2$  \cite{RabiQPT2015}.

\begin{figure}
\begin{center}
\includegraphics[width=\columnwidth]{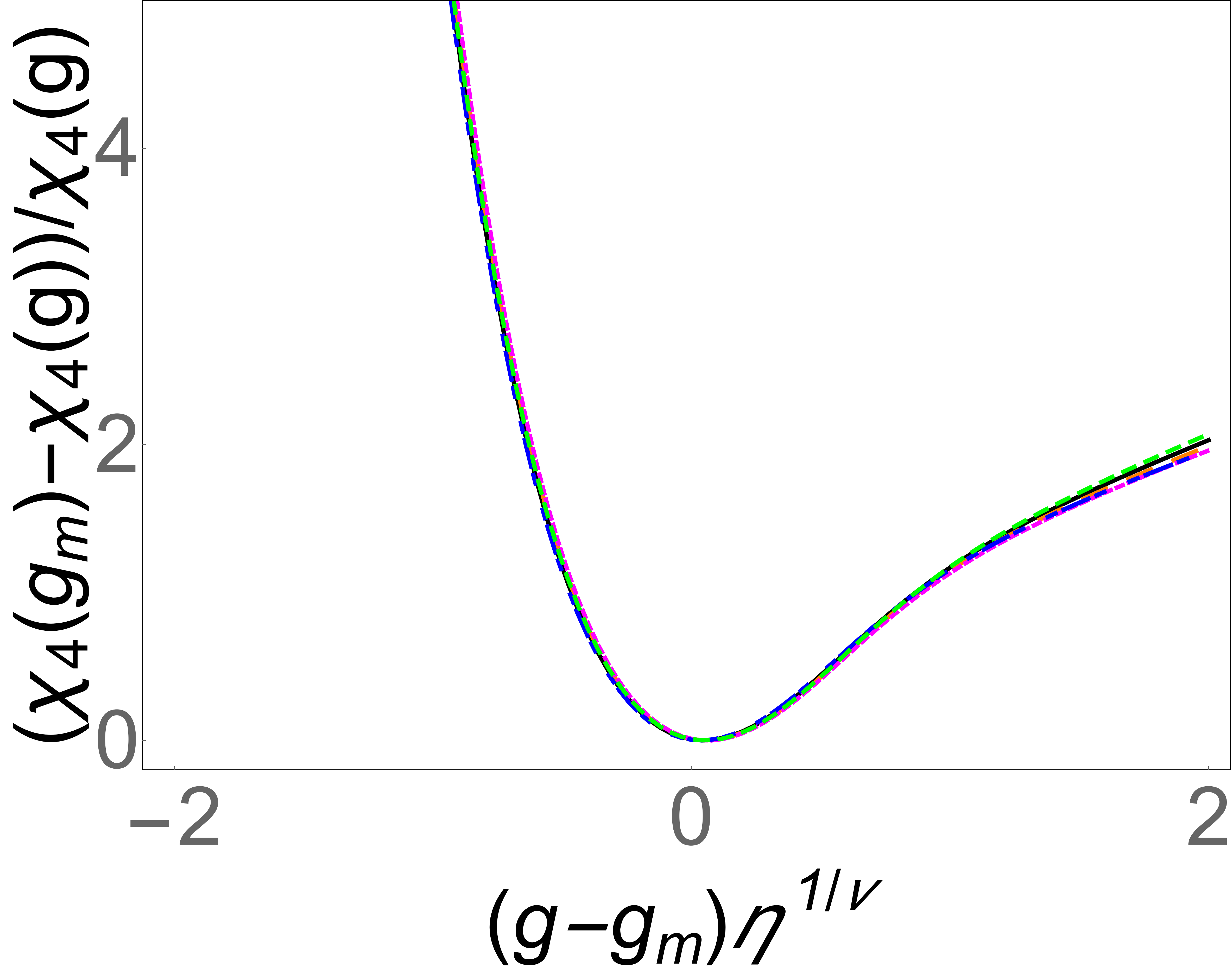}
\end{center}
\caption{(color online). Data collapse of the generalized fidelity susceptibility $\chi_4$ in the Rabi model shown in Figure 4. $\chi_4$ is a function of $\eta^{1/\nu}(g-g_m)$ only with $g_m$ being the position of the maximum of the generalized fidelity susceptibility and $\nu=1.49$ being chosen so that data for different $\eta$ collapse perfectly.}
\label{fig:epsart6}
\end{figure}

\section{Experimental consideration}
The fidelity susceptibility and generalized adiabatic susceptibility are important theoretically because they reveal the universality classes of quantum phase transitions with no knowledge of the symmetry of the system is required. However, how to detect the fidelity susceptibility and the generalized adiabatic susceptibility experimentally? It was shown that the fidelity susceptibility is related to the quantum noise spectrum of the driven Hamiltonian \cite{You2015}. Let us consider a generalized Hamiltonian $H=H_0+gH_1$ with $g$ being the control parameter. The quantum noise spectrum of the driven Hamiltonian $H_1$ can be defined as
\begin{eqnarray}
S_Q(\omega)&=&\sum_{n\neq0}\left|\langle\Psi_n|H_1|\Psi_0\rangle\right|^2\delta(\omega-E_n+E_0),
\end{eqnarray}
where $|\Psi_n\rangle$ is the eigen state of the Hamiltonian $H(g)$ with the corresponding eigen energy $E_n$, i.e. $H|\Psi_n\rangle=E_n|\Psi_n\rangle$. Note that the quantum noise only take into account the states which are different from the ground state. With such a definition for the quantum noise spectrum, the fidelity susceptibility can be written as
\begin{eqnarray}
\chi_F&=&\int_{-\infty}^{\infty}d\omega\frac{S_Q(\omega)}{\omega^2}.
\end{eqnarray}
This means that the fidelity susceptibility is the minus second moment of the quantum noise spectrum \cite{You2015}. Similarly one can show that the generalized adiabatic susceptibility $\chi_{2r+2}$ is related to the quantum noise spectrum by
\begin{eqnarray}
\chi_{2r+2}&=&\int_{-\infty}^{\infty}d\omega\frac{S_Q(\omega)}{\omega^{2r+2}}.
\end{eqnarray}

It is well known that the quantum noise spectrum can be experimentally measurable through the experiments in linear response regime \cite{fd1966}. Thus the fidelity susceptibility and the generalized adiabatic susceptibility could be experimentally measurable. In particular the simple structure of the quantum Rabi model pave the way for the experimental observation of the universal scaling behavior of the fidelity susceptibility.

\section{Summary}
In summary, we have investigated the quantum phase transitions in a very simple system with finite degrees of freedom, the Rabi model, from fidelity susceptibility approach.  We found that the fidelity susceptibility and the generalized adiabatic susceptibility present universal finite size scaling behaviors near the quantum critical point of the Rabi model if the ratio $\eta$ between frequency of the two level atom and frequency of the cavity field is finite. From the finite size scaling analysis of the fidelity susceptibility and the generalized adiabatic susceptibility, we found that the adiabatic dimension of the fidelity susceptibility $\chi_F$ and the generalized adiabatic susceptibility $\chi_4$ in the Rabi model are $4/3$ and $2$, respectively. Furthermore the correlation length critical exponent and the dynamical critical exponent at the quantum critical point of Rabi model are found to be $3/2$ and $1/3$ respectively. Since the fidelity susceptibility and the generalized adiabatic susceptibility are the moments of the quantum noise spectrum which is directly measurable by experiments in linear response regime, the scaling behavior of the fidelity susceptibility in the Rabi model could be tested experimentally. The simple structure of the quantum Rabi model provides a practical platform for experimentally verifying the scaling behavior of the fidelity susceptibility and the generalized adiabatic susceptibility at a quantum phase transition.

\begin{acknowledgements}
This work was supported by the National Natural Science Foundation of China (Grant Number 11604220) and the Startup Fund of Shenzhen University under (Grant number 2016018).
\end{acknowledgements}

\end{document}